\documentclass{emulateapj}

\usepackage{lscape} 

\def\erg/cm2sec{ergs~cm$^{-2}$~s$^{-1}$}  
\def\ergcm2{ergs~cm$^{-2}$}

\def\Chandra{${\it Chandra}$\ }
\def\HST{${\it HST}$\ }

\newcommand{\Msun}{\ifmmode {M_{\odot}}\else${M_{\odot}}$\fi}
\newcommand{\Lsun}{\ifmmode {L_{\odot}}\else${L_{\odot}}$\fi}
\newcommand{\Rsun}{\ifmmode {R_{\odot}}\else${R_{\odot}}$\fi}

\shorttitle{New Quiescent LMXBs in 47 Tuc}
\shortauthors{Heinke et al.}

\begin{document}
\title{Three Additional Quiescent Low-Mass X-ray Binary Candidates in 47 Tucanae}   

\author{C.~O.~Heinke\altaffilmark{1,2,3},
  J.~E.~Grindlay\altaffilmark{1}, P.~D.~Edmonds\altaffilmark{1}} 
\altaffiltext{1}{Harvard-Smithsonian Center for Astrophysics,
60 Garden Street, Cambridge, MA  02138;\\
 cheinke@cfa.harvard.edu, josh@cfa.harvard.edu, pedmonds@cfa.harvard.edu}

\altaffiltext{2}{Northwestern University, Dept. of Physics \&
  Astronomy, 2145 Sheridan Rd., Evanston, IL 60208}

\altaffiltext{3}{Lindheimer Postdoctoral Fellow}


\begin{abstract}
We identify through their X-ray spectra one certain (W37) and two probable
(W17 and X4) quiescent low-mass X-ray binaries (qLMXBs) containing 
neutron stars in a long \Chandra X-ray exposure of the globular
cluster 47 Tucanae, in addition to the two previously known qLMXBs.  
W37's spectrum is dominated by a blackbody-like component consistent
with radiation from the hydrogen atmosphere of a 10 km neutron star. 
W37's lightcurve shows strong X-ray variability which we attribute to
variations in 
its absorbing column depth, and eclipses with a probable 3.087 hour period. For
most of our exposures, W37's blackbody-like emission (assumed to be from the
neutron star surface) is almost completely obscured, yet some soft
X-rays (of uncertain origin) remain. 
Two additional candidates, W17 and X4, present  X-ray spectra dominated
by a harder 
component, fit by a power-law of photon index $\sim$1.6-3. 
An additional soft component is required for both W17 and X4, which
can be fit with a 10 km hydrogen-atmosphere neutron star model. 
   X4 shows significant variability, which
may arise from either its power-law or hydrogen-atmosphere
spectral component.  Both W17 and X4 
show rather low X-ray luminosities, $L_X$(0.5-10 keV)
$\sim5\times10^{31}$ ergs 
s$^{-1}$.  All three candidate qLMXBs would be 
difficult to identify in other globular clusters, suggesting an
additional reservoir of fainter qLMXBs in globular clusters that may be of
similar numbers as the group of previously identified objects.  The
number of millisecond pulsars inferred to exist in 47 Tuc is less
than 10 times larger than the number of qLMXBs in 47 Tuc, indicating
that for typical inferred lifetimes of 10 and 1 Gyr respectively, 
their birthrates are comparable. 
\end{abstract}


\maketitle

\section{Introduction}\label{s:intro}

Several soft X-ray transients, identified in outburst to be accreting
 neutron star (NS) 
 systems, have been observed in quiescence \citep[see][]{Campana98a,
 Rutledge02b}.   These systems show  
soft spectra, generally consisting of a thermal,
blackbody-like component, and in most cases a harder component extending to
higher energies, usually fit with a power-law of photon index 1-2.
 The nature of their X-ray emission remains unsolved.   \citet[see
 also Campana et al. 1998]{Brown98}  
advanced the idea that the soft thermal component seen in these 
systems can be explained by the release, over long timescales, of heat
injected into the deep crust by pycnonuclear reactions driven during
accretion (the ``deep crustal heating'' model).  This scenario 
accurately predicts the quiescent luminosity of some qLMXBs, based on their
outburst history \citep{Rutledge00,Rutledge01b}, and
 must operate at some level.  (Recent observations of low quiescent
 luminosities in some qLMXBs 
 \citep{Colpi01,Campana02,Yakovlev03,Wijnands04} 
 may require nonstandard cooling processes.) 

However, the deep crustal heating model cannot explain the hard power-law
component, which is attributed to continued accretion and/or a
shock from a pulsar wind \citep[see][]{Campana98a,Menou01}.  
Continued accretion has also been suggested as an explanation for the
thermal component, as the radiation spectrum from matter accreting radially
onto a neutron star should be similar to that from the ionized
hydrogen atmosphere of a hot non-accreting neutron star
\citep{Zampieri95}.  
Models without any continuing accretion have difficulty explaining the 
short-timescale 
($\sim10^4$ s) variability observed from Aquila X-1 and Cen X-4 
\citep{Rutledge02b, Campana03}, and the days-timescale variability 
observed from Cen X-4
\citep{Campana97} and other qLMXBs.  However, many qLMXBs in globular
clusters show neither significant variability nor an additional
power-law spectral component \citep[hereafter HGL03]{Heinke03a}.
Understanding the 
emission process will be critical for modeling the observed spectrum
and deriving fundamental physical parameters, such as mass, radius,
and magnetic field \citep{Lattimer04,Brown98}.

  Globular clusters are
 overabundant in accreting NS systems compared to the field, with
 thirty-eight probable 
qLMXBs and active LMXBs known so far \citep{Heinke03d}.  The
 well-studied globular cluster 47 
Tucanae (NGC 104; hereafter 47 Tuc) is ideal for X-ray studies of its
 binary populations \citep{Grindlay01a} due to its
close distance \citep[4.85$\pm0.18$ kpc, ][]{Gratton03} and low
reddening \citep[$E(B-V)=0.024\pm0.004$, ][]{Gratton03}.  Two qLMXBs
in 47 Tuc,
X5 and X7 \citep[originally detected by ROSAT,][]{Verbunt98},
have been spectrally identified by HGL03. HGL03 constrained the range
 of mass and radius space for X7 
using several different possible assumptions about its spectrum, and
 showed that X5 is an eclipsing system with an 8.67 hour period and
 strong dipping activity.  

We have shown \citep[HGL03,][]{Heinke03d} that the identified qLMXBs in 47
Tuc, and in globular clusters generally, have little to no hard
power-law component in contrast to the best-studied field systems,
Cen X-4 and Aql X-1 \citep{Rutledge01b, Rutledge02b}.
The exceptions thus far have been the transient LMXB in NGC 6440, which
entered outburst 13 months after the \Chandra\ observation
\citep{intZand01}, and the transient LMXB in Terzan 5
\citep{Wijnands03b}, both relatively bright qLMXBs.  
If the strength of the power-law component (and short-term variability) 
measures continuing low-level
accretion \citep{Campana98a,Rutledge02a}, then the absence of these 
indicators in globular cluster systems might be taken to indicate 
extremely low levels of accretion 
activity.  Recently, \citet{Jonker04} noted that the relative
strength of the spectral component fit with power-laws in quiescent NS
transients seems to increase for X-ray luminosities significantly
larger or smaller than $L_X\sim10^{33}$ ergs s$^{-1}$, where its
relative strength is often at a minimum \citep[but cf.][]{Wijnands03b}.
 They suggested that this spectral component arises from
different processes (accretion vs.~pulsar wind shock?) in systems with
X-ray luminosities above or below $10^{33}$ ergs s$^{-1}$. 

In late 2002, we obtained deeper \Chandra observations of 47 Tuc
\citep{Heinke04a}, allowing detailed spectral modeling of X-ray sources
as faint as $L_X\sim10^{31}$ ergs s$^{-1}$.  Most sources were found
to be consistent with powerlaw or thermal plasma models, but some
sources required more complicated models.   The highly variable X-ray
source W37 (=CXOGLB J002404.9-720451) showed a steep spectrum well-described
by a blackbody, and the X-ray sources W17 (=CXOGLB J002408.3-720431) and
X4 (=W125, or CXOGLB J002353.9-720350) showed two-component spectra in which 
the soft component could be well-represented by a model of a hydrogen
atmosphere appropriate for a neutron star.   X4 was first identified in 
ROSAT HRI data \citep{Verbunt98}, while W17 and W37 are located in
areas too crowded to 
have been identified in lower-resolution ROSAT data at their current fluxes. 
None of these three
have been positively identified with optical counterparts of any
category \citep{Edmonds03a, Edmonds03b}.  In this paper, we study
these three strong qLMXB candidates in detail. 
We describe the observations and our reduction in \S~\ref{s:obs},
the timing analysis in \S~\ref{s:timing},
and the spectral analyses in \S~\ref{s:spec}.  Discussion and conclusions are
in \S~\ref{s:disc}.

\section{Observations and Analysis}\label{s:obs}

The data used in this paper are from the 2000 and 2002
\Chandra  
observations of the globular cluster 47 Tuc.  Both sets of
observations and their initial reduction are described in
detail in \citet{Heinke04a}; prior analyses of the 2000 dataset
are described in \citet{Grindlay01a,Grindlay02} and HGL03.  Both
observations interleaved short exposures 
using subarrays (chosen to reduce pileup, which is not a significant
issue for the three sources in this paper) with longer exposures. The
observations are summarized in Table~\ref{tab:obs}; we note that
OBS\_ID 3385 
suffered strongly increased background, so we do not use that short 
observation.   We reprocessed (using CIAO 3.0) both the 2000
and 2002 observations to remove the 0\farcs25 pixel 
randomization added in standard processing, implement the CTI
correction algorithm on the 2000 data, and use updated (Feb. 2004,
CALDB v. 2.26)
calibration files.  We did not remove events 
flagged as cosmic-ray afterglows by the standard processing.  
290 sources
were detected using the WAVDETECT routine in the 0.3-6 and 0.3-2 keV
energy bands, and ten additional sources (missed by WAVDETECT due to 
crowding) were added to the source list by hand.

We used the ACIS\_EXTRACT routine \citep{Broos02} from the Tools for
X-ray Analysis
(TARA\footnote{http://www.astro.psu.edu/xray/docs/TARA/}) to construct
 polygonal extraction regions generally chosen to match the 90\%
encircled energy (at 1.5 keV) from the \Chandra  point-spread
function (PSF).  
We extracted source spectra and background spectra from nearby 
source-free regions, and corrected the ancillary response function
(ARF) for the energy-varying fraction of the PSF enclosed by the
extraction region.  We used the Jan.~2004 release of the ACIS
contamination model to correct for the hydrocarbon buildup on the
detectors \citep{Marshall04}.  We adjusted all event times to the
solar system barycenter using 
satellite orbit files provided by the \Chandra X-ray Center.  

\subsection{Timing Analysis}\label{s:timing}

Kolmogorov-Smirnov (KS) tests on the 2002 event arrival times from W37
and X4 
confirm that both are clearly variable on short
timescales, with KS probabilities of constancy reaching $10^{-47}$ and
$3\times10^{-4}$ for the first 2002 observations of W37 and X4
respectively.   W17 shows only marginal evidence for variability, with
KS probabilities reaching only 0.04.  X4 falls on a chip gap in the
2000 data, and W37 is much fainter in the 2000 data.  W37 receives a
photon flux over five
times higher in the first 2002 observation than in any other
observation, and 75\% of the 2002 photons from W37 arrive in the first
quarter of the data.  X4 appears to have dimmed slightly between the first
and fourth long 2002 observations (at 99\% conf.). 

\begin{figure}
\figurenum{1}
\epsscale{0.8}
\plotone{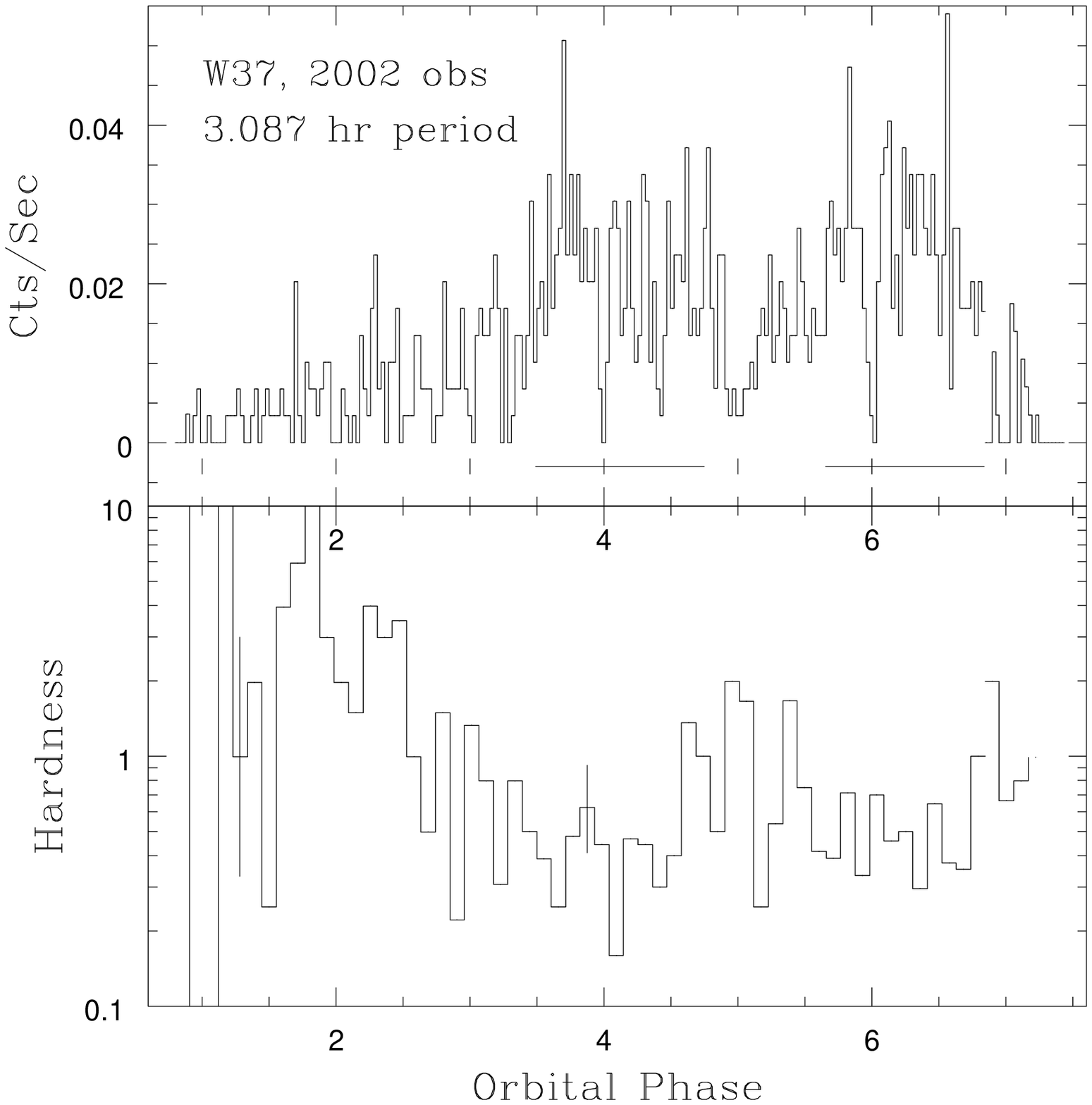}
\caption{Top: Lightcurve of W37 (0.3-8 keV), in 300 second
bins, from first two 2002 observations (OBS\_IDs 2735 and 3384, 
65 and 5 ksec, respectively). Time is
labeled in units of the best-fit period, 11112.5 
seconds, with integer phases occurring at the times of mid-eclipse.
Horizontal bars indicate portions of the data taken for a high-state
spectrum (2735,H); the remainder of OBS\_ID 2735 is taken as a 
low-state spectrum (2735,L).  
Bottom: Hardness ratio (1-6 keV/0.3-1 keV) lightcurve for
W37, corresponding to the same timespan as above, but with 1200 second
bins.  An anticorrelation can be seen between hardness and count
rate.}\label{fig:w37lc}
\end{figure}

\begin{figure}
\figurenum{2}
\epsscale{0.8}
\plotone{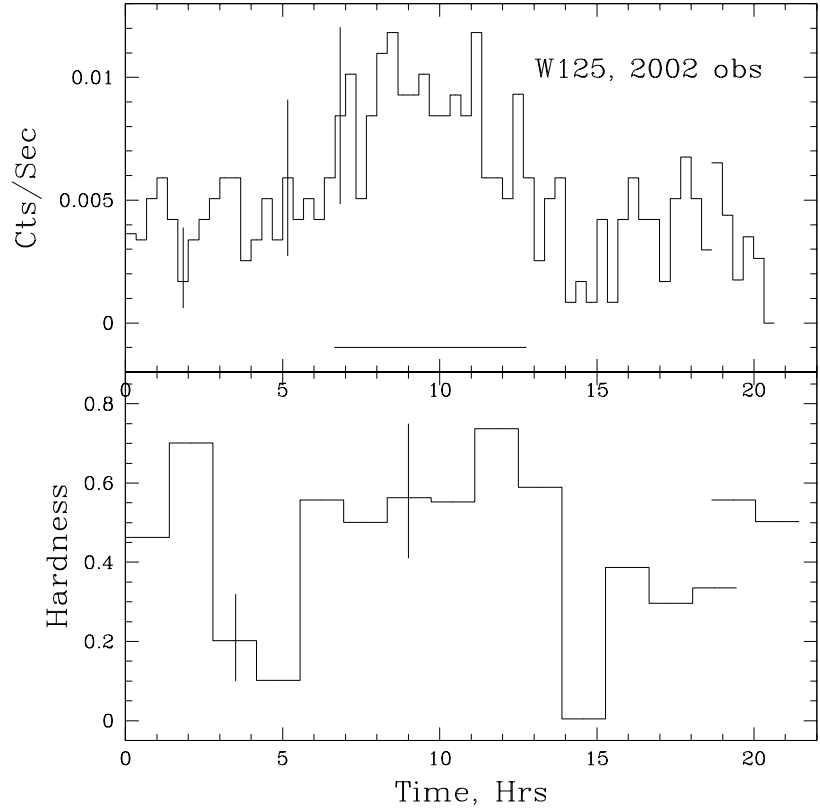}
\caption{Top: 0.3-8 keV lightcurve of X4 (W125), in
1200 second bins, from first two 2002 observations.  Time is labeled
in hours.  A horizontal bar marks the portion of the data taken for
the high-state spectrum.  Bottom: Hardness ratio (1-6 keV/0.3-1 keV)
lightcurve for X4, with 5000 second bins.  A correlation can be seen
between hardness and count rate. } \label{fig:x4lc}
\end{figure}

\begin{figure}
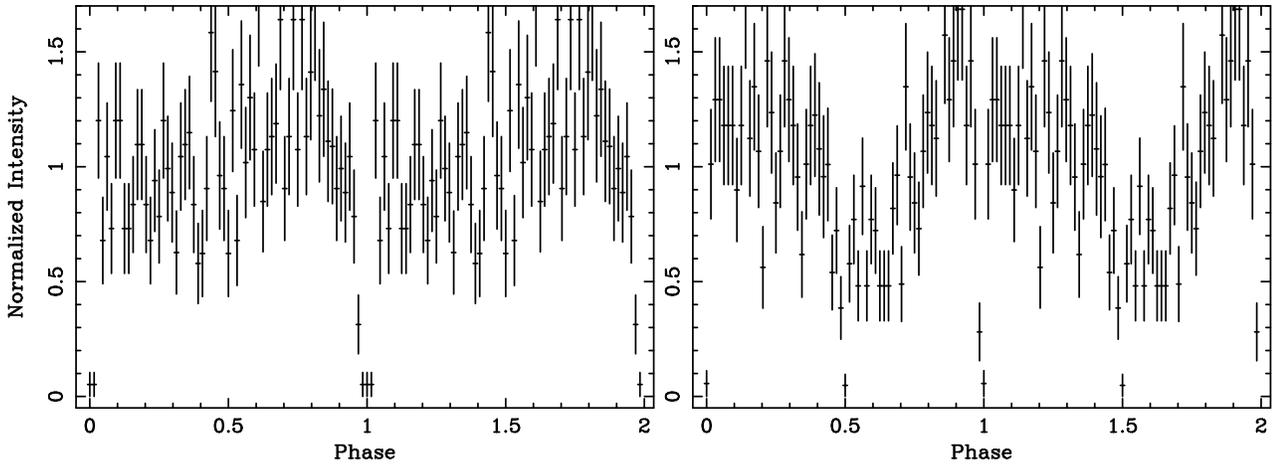

\figurenum{3}
\epsscale{0.7}
\includegraphics[angle=270,scale=.36]{f3a.eps}
\includegraphics[angle=270,scale=.36]{f3b.eps}
\caption{Left: Folded lightcurve of all W37 2002
 data on 11112.5 second period, with the data repeated over two phases
 for clarity.  A clear eclipse can be seen at phase 0.  Each bin is
 185.2 seconds long.  Right: Folded lightcurve of all W37 2002 data on 
22225 second period, showing eclipses at phases 0 and 0.5. 
}\label{fig:w37fold}
\end{figure}

We show the
first (and most variable) sections of the 2002
lightcurves for W37 and X4 
in Figures~\ref{fig:w37lc} and \ref{fig:x4lc}.  Generating a power
spectrum for W37 (using XRONOS,
\footnote{http://heasarc.gsfc.nasa.gov/docs/xanadu/xronos/xronos.html})
gives a possible peak around 11000 seconds along with strong red
noise. An  epoch folding search using all the 2002 observations finds
a periodicity at 11112.5 ($\pm15$)  
seconds.  Careful inspection of W37's lightcurves reveals 
likely eclipses, indicated in Figures~\ref{fig:w37lc} and
\ref{fig:w37fold} (left; a folded lightcurve), consistent with the 
11112.5 second 
period.  The two clearest eclipses are separated by 6.2 hours, so we 
also include a lightcurve folded on twice our preferred period 
(Figure~\ref{fig:w37fold}, right), showing two apparent eclipses.  This 
indicates that our preferred period is probably correct, but the low
count rate of the data may allow for alternative solutions.
The folded lightcurve suggests eclipses $550\pm200$ seconds in length.

 We note
that W37's lightcurve is similar to that of X5, 
presented in HGL03 and Heinke et al.~(2005, in prep), suggesting that
the strong variability seen in W37 is also probably due to a varying column
of cold gas associated with an accretion disk viewed nearly
edge-on. However, W37 may show residual emission during eclipses,
which is not seen for X5 (HGL03).  The period indicates a companion 
mass of order 0.34 \Msun, if the companion has a density appropriate 
for lower main sequence stars \citep[][eq. 4.9]{Frank92}.  Since 
these companions are sometimes bloated 
\citep{King99,Podsiadlowski02,Orosz03,Kaluzny03b}, the true mass 
may be smaller.  Assuming a 1.4 \Msun\ neutron star, the orbital separation
would be $\sim9\times10^{10}$ cm.

We calculate the hardness ratio for W37 as the ratio of the counts
detected in the 1-6 keV and 0.3-1 keV bands (this choice divides the
detected counts roughly evenly), and plot this ratio in the lower
panel of Figure~\ref{fig:w37lc}.  Comparison with the upper panel shows that
decreases in the observed flux correlate well with
increasing hardness. We confirm this with a Spearman rank-order
correlation test \citep{Press92} comparing the binned (1200 second)
lightcurve with the similarly binned hardness ratio.  This gives
$r_s$=-0.446, a negative correlation, with chance probability
$5.6\times10^{-4}$. 
 This is a strong sign that the variation is due to
changing photoelectric absorption.   We divide W37's long 2002
observations into high and low flux states (e.g., the high flux state
portions indicated in Figure~\ref{fig:w37lc}), which we use to extract
  spectra.    
 
X4's total and hardness ratio lightcurve for OBS\_IDs 2735 and 3384
are shown in Figure~\ref{fig:x4lc}. No 
periodicities are apparent in X4's lightcurve or power spectra.
X4's spectrum appears to harden as it brightens, in contrast to
W37, which indicates the variability is not due to increasing
photoelectric absorption.  We confirm this with a Spearman rank-order
correlation test on OBS\_ID 2735, finding $r_s$=0.444 and chance probability
$6.0\times10^{-4}$ for the binned (1200 second) lightcurve and hardness
ratio.  We extract high
and low portions of X4's lightcurve from the first observation for
spectral fitting, which are marked in Figure~\ref{fig:x4lc}.

\subsection{Spectral Analysis}\label{s:spec}

In \citet{Heinke04a}, we identify W37, X4, and W17 as
possible qLMXBs by their agreement with models consisting of a
hydrogen-atmosphere model and a power-law, along with their failure to fit
single-component models typical of thermal plasma (such as seen in
cataclysmic variables and X-ray active binaries).  Here, we analyze
time-resolved 
spectra for W37 and X4, and use the merged spectra (2000 and
2002) for W17 (since it shows no variability).  We use
the XSPEC {\it phabs} model to describe neutral absorbing gas.  We use
the \citet{Lloyd03} model to describe thermal blackbody-like emission
from the hydrogen atmosphere of a neutron star, fixing the redshift 
to 0.306 (appropriate for a 1.4 \Msun, 10 km neutron star) and take
the distance to be 4.85 kpc \citep{Gratton03}.   The total number of
counts in the 2002 spectra of W37, W17 and X4 were 1277, 909, and
1426, while from the 2000 data 46, 114, and 62 counts were available.
We binned spectra with more than 1000 counts with 40 counts/bin,
spectra from the 2000 observations or with fewer than 50 counts at 10
counts/bin, and other spectra at 20 counts/bin. 

\subsubsection{W17}\label{s:w17}

\begin{figure}
\figurenum{4}
\epsscale{0.6}
\includegraphics[angle=270,scale=.60]{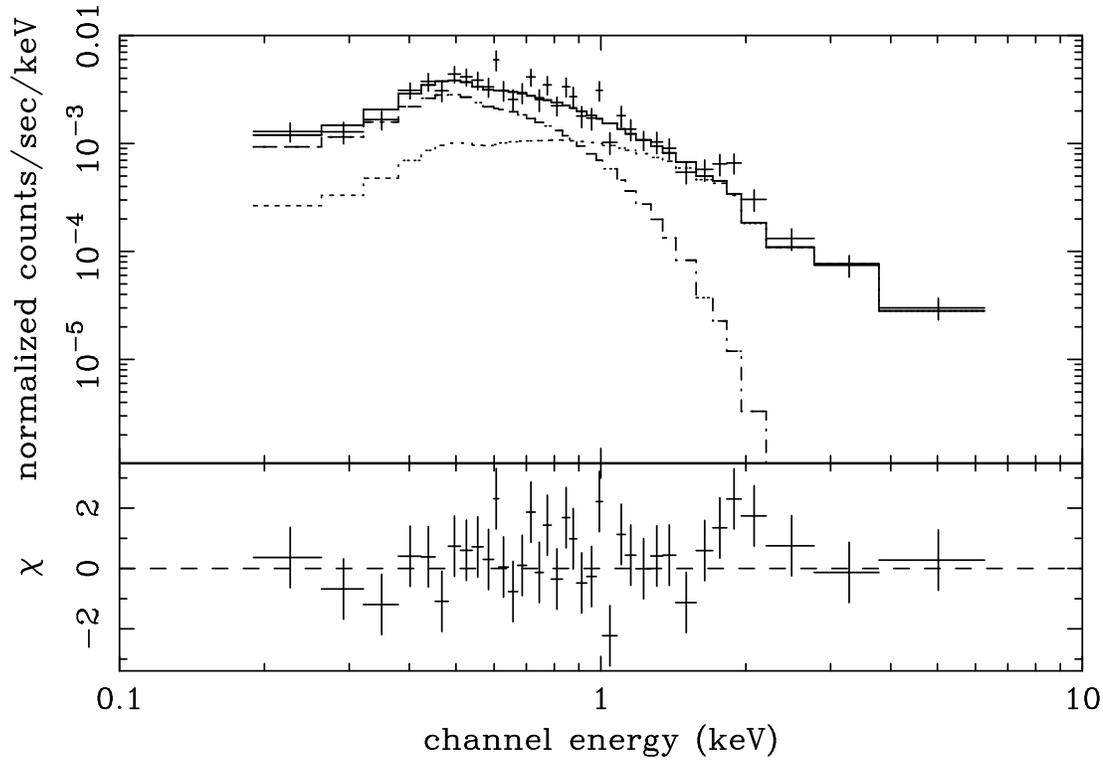}
\caption{X-ray spectrum of W17 from the 2002 observations, fit with an
  absorbed hydrogen atmosphere 
model and power-law as in Table~\ref{tab:spec}. The contributions from
  the two components are shown, with the power-law dominating at
  higher energies.}\label{fig:w17spec}
\end{figure}

W17 shows no variability on short or long timescales, so we
fit the combined 2002 and combined 2000 datasets simultaneously.  We
try a variety of absorbed single-component models including blackbody,
bremsstrahlung, MEKAL, power-law, and hydrogen atmosphere neutron star
models, but do not find good fits with any.  The best fit of these, a
power-law model, gives $\chi^2_{\nu}=1.36$, for a 2\% null hypothesis
probability (nhp).  Models consisting of two mekal components also do not
produce good fits.  Adding a soft component, such as the hydrogen-atmosphere
neutron star model of \citet{Lloyd03}, to the power-law model improves
the statistics tremendously ($\chi^2_{\nu}=1.05$, nhp=37\%; F-test
gives probability of $9\times10^{-5}$ that the extra component is not
needed).  The inferred radius of the neutron star model,
$R=11.8^{+9.7}_{-3.9}$ km, is perfectly consistent with the canonical
1.4 \Msun, 10 km neutron star predictions, although it cannot usefully
constrain the neutron star structure. (The inferred radius for a
blackbody would be only $1.2^{+0.1}_{-0.4}$ km.) A qLMXB model 
is thus an excellent explanation for this spectrum, shown in
Figure~\ref{fig:w17spec}. The parameters for a  
fit with a Lloyd H-atmosphere model and power-law are listed in 
Table~\ref{tab:spec}.   The power-law component provides a majority
($65^{+28}_{-22}$\%) of the 0.5-10 keV flux, and the thermal component is
quite weak, with unabsorbed X-ray luminosity
$L_X(0.5-2.5)=1.7\times10^{31}$ ergs s$^{-1}$ and inferred bolometric
luminosity $L_{bol}=6.7\times10^{31}$ ergs s$^{-1}$.  

\subsubsection{W37}\label{s:w37}

We identify pieces of W37's lightcurves at high and
low flux levels in the four long observations, with only a low flux
level in the second long observation; the (barycentered) times are
listed in Table~\ref{tab:add}.  We 
extracted spectra from these periods, as well as from the summed 2000
observations, and background spectra from
surrounding source-free 
regions, producing eight spectra of varied quality.  

 The hardening
of the spectrum with decreasing flux suggests that variation of the
column density is responsible for the spectral variations, and so we
allow the gas column to vary between spectral segments. However, no
absorbed single-component model was able to describe the eight   
spectra simultaneously by only varying the absorption and one 
other model parameter.  (Allowing all the parameters to vary freely allows 
some reasonable fits, but loses physical meaning.  For instance, a
blackbody with strongly varying normalization, temperature and
absorbing column, while a reasonable fit, is difficult to interpret.)  
No single- or double-temperature thermal plasma model
(represented by VMEKAL models in XSPEC, set to cluster abundances)
provides a good fit to the brightest portions of the data.  The brightest 
portions of the data are best fit by a blackbody or hydrogen
atmosphere neutron star model with varying absorption. If described by
a blackbody, the inferred radius is only 
$1.4\pm0.2$ km in size.  This component is well-described by a 
hydrogen-atmosphere neutron star model with a constant 
physical radius of order 10 km
($12.3^{+5.8}_{-3.5}$ km) and constant temperature ($82^{+10}_{-9}$ eV).  
Therefore we infer that W37 is a qLMXB containing a neutron star.

\begin{figure}
\figurenum{5}
\includegraphics[angle=270,scale=.60]{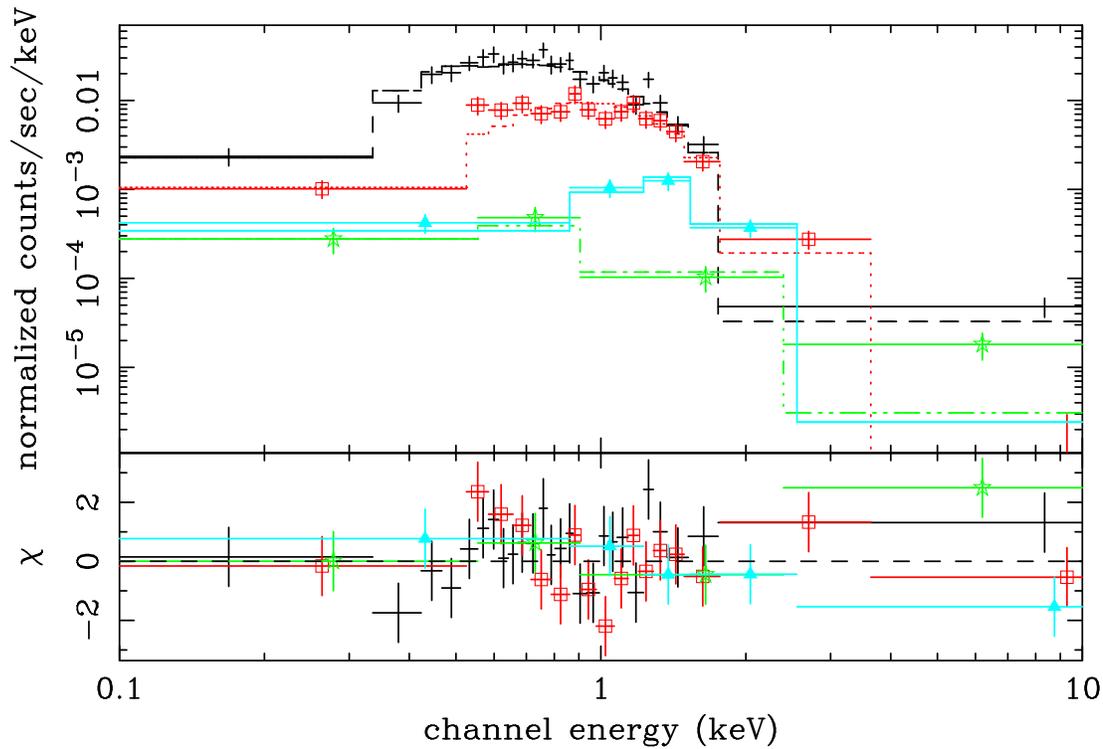}
\caption{X-ray spectra of W37, taken from four different
portions of the data with different fluxes.  The spectra were fit with
an absorbed hydrogen atmosphere model plus a separately absorbed
power-law.  Only the absorption on the hydrogen-atmosphere model was
allowed to change between data segments.  Eight portions were fit
to produce the results in Tables~\ref{tab:spec} and ~\ref{tab:add},
but only four are shown here to 
reduce confusion.  From highest to lowest flux, these are [2735,H],
dashed line; [2735,L], squares and dotted (red) line; [2737,L],
triangles and solid (blue) line;
[2736], stars and dash-dotted (green) line. (See the electronic edition of the
Journal for a color version of this figure.)}\label{fig:w37spec}
\end{figure}

 The fainter
portions show additional soft flux beyond what can be accounted for by
an absorbed H-atmosphere (or blackbody) component.  This suggests a
second component that does not suffer the same absorption as the
primary spectral component.  We model this component with a separately
absorbed power-law, although a variety of models can account for this
low-count component equally well (including a 0.6 keV bremsstrahlung
spectrum, a 
0.1 keV blackbody with inferred 1 km radius, or a 0.13 keV thermal
plasma model), and we do not know its physical origin.  
We hold the faint  
component, its absorption, and the H-atmosphere model parameters fixed 
between data portions, and
vary only the absorption to the H-atmosphere model.  This is the
XSPEC model PHABS ( POW + PHABS * HATM ), varying only the second PHABS
component between observations.   We find a good
fit ($\chi^2_{\nu}=1.24$ for 57 degrees of freedom, or dof, and an 
nhp of 10.6\%).  We are unable to impose
serious constraints upon this model if additional parameters are
allowed to vary, since the faint portions of the data have very few
counts. Therefore we cannot usefully constrain any temporal or 
spectral variability in the faint component.  
We show part of this fit (only 4
of 8 spectra to reduce confusion) in Figure~\ref{fig:w37spec}, and
give details of the fit in 
Tables~\ref{tab:spec} and \ref{tab:add}.   The fitted absorption column 
to the probable neutron star varies by a factor $>100$, decreasing to 
a value consistent with the cluster column during parts of the
first 2002 observation. The faint component
is responsible for a very small portion ($2^{+17}_{-1}$\%) of the
unabsorbed flux, and produces a photon index of $3^{+5}_{-1}$. 
The absorption column to the faint 
component is poorly constrained ($N_H=4.5^{+84}_{-4.5}\times10^{20}$) 
but consistent with the column to the cluster.  The
lowest-flux spectrum (from OBS\_ID 2736), dominated by the faint 
component, is one of the spectra shown in Figure~\ref{fig:w37spec}.

\subsubsection{X4}\label{s:x4}

X4 varies during the first 2002 observation, but is not clearly
variable within other observations.  We separately fit the high-flux
and low-flux parts of the first 2002 observation (OBS\_ID 2735), and
data from each of the other three long 2002 observations, plus one
spectrum from the combined 2000 
observations.  We found that the remaining 2002 observations were
spectrally indistinguishable, so we combined all 2002 data except
OBS\_ID 2735 into one spectrum, and fit the high and low-flux parts of
2735, the rest of the 2002 data, and the combined 2000 data
simultaneously.  Fits with any single absorbed component, with all
parameters allowed to vary, failed to give reasonable fits to the
spectra.  A power-law fit gave the lowest $\chi^2_{\nu}=1.45$,
nhp=3.9\%, if $N_H$ is allowed to vary substantially along with the
power-law parameters.  However, if $N_H$ is not allowed to vary during
the 2002 data (as indicated in \S~\ref{s:timing} above), then
$\chi^2_{\nu}$=1.52, nhp=1.9\%.  Adding a H-atmosphere model
improves the fit substantially (F-statistic=4.38, prob. 1.9\% extra
component not required), and results in an inferred radius consistent
with the 10 km canonical NS radius.  We conclude that X4 is a good
candidate for a qLMXB.  Alternate models involving two VMEKAL 
components with all temperatures and normalizations free do not
produce good fits.

\begin{figure}
\figurenum{6}
\epsscale{0.6}
\includegraphics[angle=270,scale=.60]{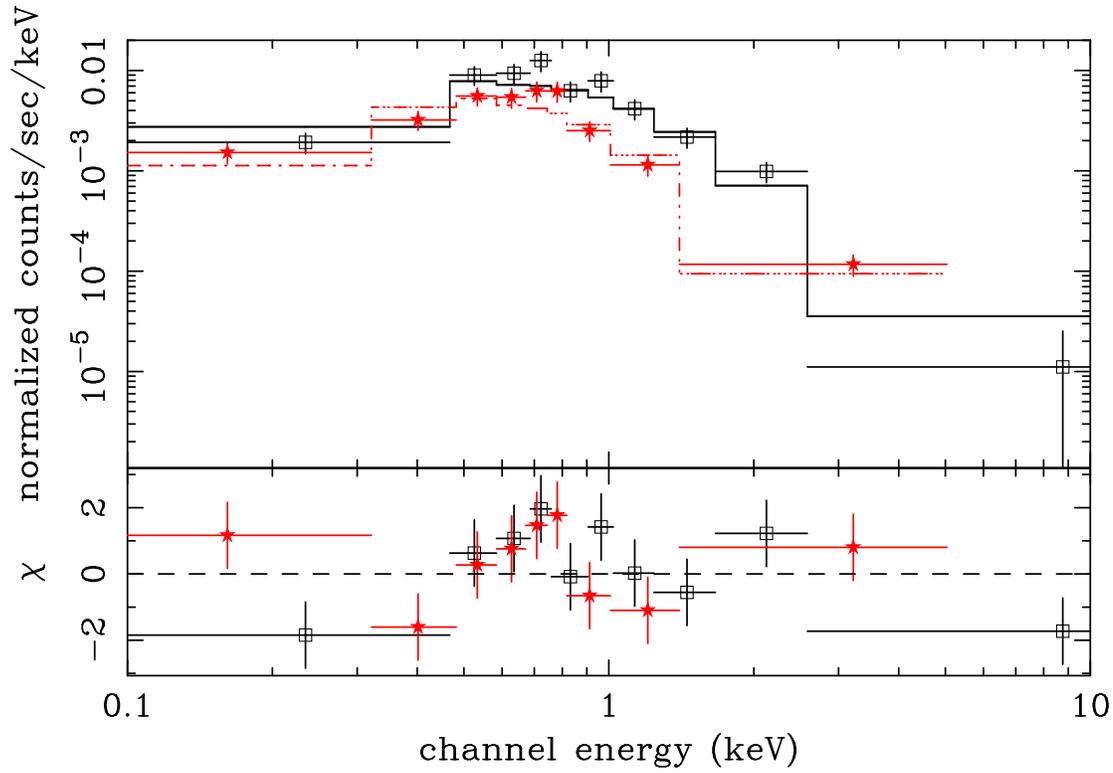}
\caption{X-ray spectra of X4 from two parts of the 
first 2002 observation, fit with an absorbed
hydrogen atmosphere model and a power-law (with the power-law
component allowed to vary between observations) as in 
Table~\ref{tab:spec}.  The lower-flux spectrum (red) shows the
greatest differences at higher energies, indicating the difference is
not due to photoelectric absorption. (See the electronic edition of the
Journal for a color version of this figure.)}\label{fig:x4spec1}
\end{figure}

\begin{figure}
\figurenum{7}
\epsscale{0.6}
\includegraphics[angle=270,scale=.60]{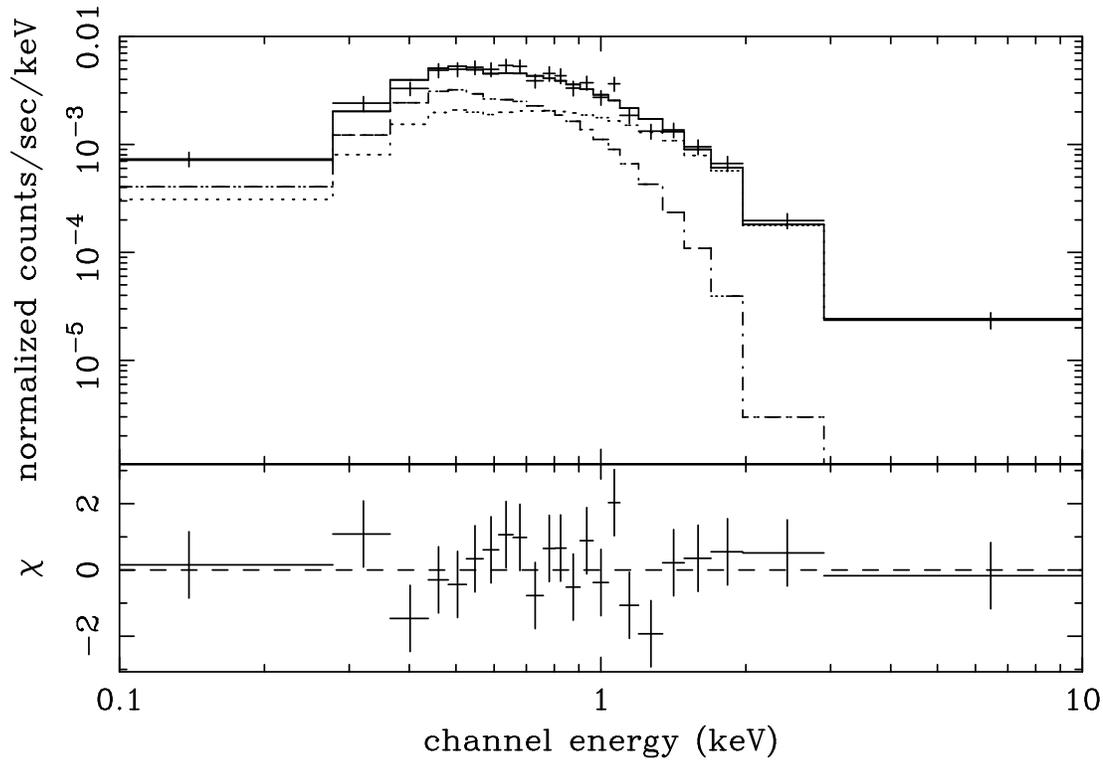}
\caption{X-ray spectra of X4 from the rest of the 2002
observations (except those in Fig.~\ref{fig:x4spec1}), fit with an absorbed
hydrogen atmosphere model and a power-law (with the power-law
component allowed to vary between observations) as in 
Table~\ref{tab:spec}.  The contributions from the two components are
shown, with the power-law dominating at higher energies.}\label{fig:x4spec2}
\end{figure}

The variability during OBS\_ID 2735 demands that at least one
component of the spectrum varies. Varying $N_H$ alone does not produce
a good fit ($\chi^2_{\nu}=2.41$, nhp=$1\times10^{-4}$\%).  Varying the
power-law component does produce a 
good fit ($\chi^2_{\nu}=1.30$, nhp=11\%; see  Table~\ref{tab:spec}).  
We show this fit for the two components of OBS\_ID 2735
(Fig.~\ref{fig:x4spec1}) and for the remainder of the 2002
observations, indicating the contributions of the two components
(Fig.~\ref{fig:x4spec2}).  Varying only the 
H-atmosphere component's temperature and radius also produces a good fit 
($\chi^2_{\nu}=1.35$, nhp=7.3\%; see Table~\ref{tab:spec}).  
Varying both the
power-law and H-atmosphere parameters ($\chi^2_{\nu}=1.21$, nhp=20\%)
is only a marginal improvement over varying a single component, and
requires correlated swings in numerous parameters (which we assume is
less likely than one or two varying parameters). Varying
only the power-law normalization also produces a good fit
($\chi^2_{\nu}=1.33$, nhp=7.5\%), with a power-law photon index of
2.2$^{+0.3}_{-0.3}$.  Varying the H-atmosphere 
temperature alone does not produce a good fit ($\chi^2_{\nu}=1.62$,
nhp=0.7\%), nor does varying the normalization (and inferred radius)
alone ($\chi^2_{\nu}=1.85$, nhp=0.07\%).
If the thermal component is due to continued accretion, and the
variation is due to changes in the accretion rate onto the 
neutron star, we would expect only the temperature to vary, not the
inferred radius.  It is possible that the emitting area varies if
accretion is ongoing, but variation in the emitting area has not been
clearly seen in field qLMXBs with higher statistics
\citep{Rutledge02b,Campana03}.   
Therefore we conclude that the power-law probably varies, and that we
do not have strong evidence for variability of the thermal component in this
source.  

We can also compare
X4's flux in our \Chandra observations to its flux in ROSAT HRI 
observations from 1992-1996 (which contain very little spectral
information).  \citet{Verbunt98} report 49$\pm9$ counts 
from X4 (which we confidently identify with our source) in 58820 seconds of
exposure.  Using a power-law spectral fit to only the 0.5-2.5 keV \Chandra
data (with $N_H$ fixed to the cluster value, $1.3\times10^{20}$
cm$^{-2}$), we find a power-law index of 2.9$\pm0.1$, with 
(absorbed) 0.5-2.5 keV X-ray fluxes
ranging between $2.7\times10^{-14}$ to $1.0\times10^{-14}$ ergs
s$^{-1}$ cm$^{-2}$ (for different parts of the \Chandra data, as
above).  Using 
this spectrum, the PIMMS tool\footnote{Available at
  http://asc.harvard.edu/toolkit/pimms.jsp} gives an absorbed 0.5-2.5
keV X-ray flux of $2.8\times10^{-14}$ ergs s$^{-1}$ cm$^{-2}$ for the
ROSAT HRI data. This flux is consistent with the upper end of the range
of fluxes observed with \Chandra.  

\section{Discussion}\label{s:disc}

The 47 Tuc X-ray sources W37, W17 and X4 are very probably qLMXBs.  No
other identified source types  
observed in globular clusters show spectral components consistent with
the emission from a 10 km radius hydrogen atmosphere, as observed in
these objects.  Millisecond pulsars show thermal emission from a much
smaller area, and at lower luminosities
\citep{Zavlin02,Grindlay02,Bogdanov04}.  Magnetic
CVs show soft, blackbody-like components plus harder emission, but
with emission radii that 
are much larger (generally hundreds of km) than the implied radii
(0.5-2 km) from blackbody fits to these sources.  We also
note that these three objects are the three X-ray brightest objects
(excepting X7, another known qLMXB) not yet
positively identified in our \HST optical identification program
\citep{Edmonds03a, Edmonds03b}.  If W17 were a cataclysmic variable
(or, even less likely, an X-ray active binary), its faint upper
limit of $U\gtrsim24$ would produce an unusual X-ray to optical flux
ratio, which led \citet{Edmonds03a} to 
suggest this object as a possible qLMXB. X4 fell outside the \HST 
fields of view analyzed by \citet{Edmonds03a}, but its X-ray spectral
similarities to W17 are convincing.  W37's error circle showed no
blue or variable objects, but two bright main-sequence stars fall
within a 2$\sigma$ error circle (not unusual considering its projected
location at the center of the cluster), likely obscuring the true
(fainter) companion \citep[as for X7,][]{Edmonds02a}.  The far-UV imaging
observations of 
\citet{Knigge02} include W37's position, and may be able to provide stronger
constraints on a possible UV counterpart.  We conclude that W37's varying
$N_H$ and 
eclipses indicate that it is a transiently accreting qLMXB, with
marked similarities to X5 (HGL03).  We also note that W37 is the
second qLMXB without a recorded outburst to have an identified period,
after X5, and that its short period implies a very dim low-mass companion. 

We can not rule out possibilities other than qLMXBs for W17 and X4.
It is possible that
W17 and X4 are neutron stars without Roche-lobe filling low-mass
companions, accreting from the intracluster medium \citep{Pfahl01} or
from the normal stellar winds of low-mass stellar companions in close
orbits, in so called pre-LMXB systems  
\citep{Willems03}.  Either of these is possible, but we judge a
stellar wind model to be more likely, because neutron stars that
accrete from the intracluster medium must also be able to accrete the
higher-density wind from a low-mass star, and because all the (slowly
spinning) neutron 
stars in the cluster should display signals of accretion from the
intracluster medium if any do.  The low bolometric luminosities from
the neutron star surfaces of W17 and X4 indicate a very low level of
time-averaged mass transfer in the \citet{Brown98} model, or enhanced
neutrino cooling. Mass transfer from a pre-LMXB by a stellar wind thus
seems a reasonable possibility.  
However, similarly low quiescent luminosities and power-law dominated
spectra have recently been found for several transient LMXBs:
$L_X$(0.5-10 keV)=$5\times10^{31}$ ergs s$^{-1}$ for SAX
J1808.4-3658 in quiescence \citep{Campana02},  $L_X$(0.5-10
keV)=$9\times10^{31}$ ergs s$^{-1}$ ($d$/8.5 kpc)$^2$ for XTE
J2123-058 \citep{Tomsick04}, and $L_X$(0.5-10 keV)=$1\times10^{32}$
ergs s$^{-1}$($d$/4.9 kpc)$^2$ for SAX J1810.8-2609 
\citep{Jonker03}.  
This suggests that W17 and X4 may be normal qLMXBs.  
They seem to be generally
consistent with the anticorrelation between X-ray luminosity, and the
fraction of that luminosity seen in the power-law component, suggested
by \citet{Jonker04}. On the other hand, W37 \citep[like the qLMXB in
  NGC 6397,][]{Grindlay01b}, appears to show a smaller power-law
component than expected from the anticorrelation of \citet{Jonker04}.

The faint soft component in W37's spectrum must be
associated with W37, 
since the position of the source does not change between the brightest
and faintest intervals (as would be expected if another X-ray source
produced this component).    The X-rays might 
be produced by the rapidly rotating companion star's corona, as
suggested \citet{Bildsten00} and disproved \citep{Lasota00,Kong02} for
quiescent black hole systems.  We can 
estimate the coronal saturation luminosity  of a
companion as ${\rm log}
L_{cor}=-2.9+2 {\rm log} R_d$, where $L_{cor}$ and $R_d$ are
measured in units of the solar luminosity and radius
respectively \citep{Fleming89}.  Assuming a companion mass of 0.3
\Msun\ (appropriate for a 
lower main-sequence star in a 3-hour orbit; see \S~\ref{s:timing}), we
compute a radius 
$2.5\times10^{10}$ cm, and thus a coronal saturation luminosity 
${\rm log} L_{cor} < 29.8$ ergs s$^{-1}$, which is significantly less
than the  $5\times10^{30}$ ergs s$^{-1}$ required for the additional
component.  Scattering of W37's X-rays in an 
accretion disk wind might produce a similar spectrum, as observed in
the eclipsing CVs OY Car, DQ Her and UX UMa
\citep{Naylor88,Mukai03,Pratt04}.  We are not able to make predictions
about the efficiency of scattering in such a wind, but we do not see
any outstanding objections to this scenario.   

Alternatively, this faint less-absorbed component might be identified
with the power-law component observed 
in many field qLMXBs \citep[e.g., ][]{Campana98a, Rutledge01a,
Rutledge01b}, and in W17 and X4.  If this identification is correct,
then this  
component must be generated in an extended environment around the
neutron star, and not near the surface.  A rough estimate of its size
can be estimated by assuming that the place of its generation is large
compared to the size of variations in the accretion disk. (This assumes 
that the variations in $N_H$ observed for the neutron star component are 
due to material in the accretion disk.)  The
accretion disk is probably of order $4\times10^{10}$ cm in radius, for
typical masses of the components (1.4 and 0.1 \Msun) and an accretion
disk size one-half the orbital separation.  Assuming
that variations in the accretion disk height may be up to one-tenth the
radius of the disk, this suggests that the environment that generates the
faint soft component should be larger than $4\times10^{9}$ cm.  This
is significantly larger than the light-cylinder radius for neutron
stars spinning at millisecond periods.  If the X-rays are generated
within or at the edge of the magnetosphere, the size of the emitting
region implies that a pulsar 
mechanism is operating, and that the observed X-rays may be a shock
from the pulsar wind as suggested by \citet{Campana98a}. 
The physical nature of the additional component to
W37's spectrum cannot be conclusively determined from this small
amount of data, but the possibility that we are seeing a pulsar wind
in a system with an accretion disk is intriguing. 

W17 and X4 indicate that a class of faint qLMXBs dominated by harder
nonthermal emission may exist in similar numbers as the class of
qLMXBs dominated by thermal emission from the neutron star surface
\citep{Heinke03d}.   Additional qLMXBs
with quiescent properties similar to SAX J1808.4-3658 hiding among the
low-luminosity sources in 47 Tuc cannot be ruled out, since the
neutron star atmosphere is not always clearly detectable
\citep{Campana02}.  
These particular identifications would be difficult with
shorter exposures, as have been performed for other globular 
clusters, due to
the burying of their telltale thermal components under a stronger
nonthermal component and/or a blanket of absorbing gas, and 
 their significantly lower fluxes.  However, it will be worthwhile to
look for similar objects in the nearest, best-studied clusters.
 The numbers of faint qLMXBs with $L_X<10^{32}$ ergs s$^{-1}$ may be
 similar to those of the identified qLMXBs. We note that there are few
hard qLMXBs among sources with 
$L_X=10^{32-33}$ ergs s$^{-1}$, using the results of detailed X-ray
and optical studies of 47 Tuc, NGC 6397, and NGC 6752
\citep{Edmonds03a,Edmonds03b,Grindlay01b,Pooley02b}, but a significant
population below $L_X<10^{32}$ 
ergs s$^{-1}$ cannot be ruled out.  If we extrapolate from these two
faint, nonthermally-dominated qLMXBs to the globular cluster system
\citep[using their close encounter frequencies, as in][and Poisson
  statistics]{Heinke03d}, 
an additional 56$^{+61}_{-36}$ such 
systems are suggested, added to the $\sim$100 qLMXBs already
anticipated \citep{Pooley03,Heinke03d}.  
Study of faint qLMXB systems may shed light
on the interior physics of neutron stars, the physics of accretion at
low mass transfer rates, and possibly the transition as LMXBs evolve to
become millisecond pulsars \citep{Burderi02, Burderi03}.  

Early estimates of the numbers of  millisecond
pulsars ($\sim$10000) and their suggested progenitors, LMXBs (12), in
globular clusters indicated (for LMXB lifetimes 1/10 those of
millisecond pulsars, $\sim$1 and 10 Gyr) a large discrepancy in 
their respective birthrates \citep{Kulkarni90,Hut91}, thus implying
that other formation processes may have created some globular cluster
pulsars \citep{Bailyn90a}.  
Radio timing surveys of 47 Tuc have identified 22 millisecond pulsars
\citep{Camilo00, Freire01a}, and X-ray, optical and radio 
imaging surveys indicate that the total number of millisecond pulsars
in 47 Tuc is of order 30 \citep{Heinke04a,Edmonds03b,McConnell04,Grindlay02}. 
When these constraints are compared with the current number of
likely quiescent LMXBs in 47 Tuc (5), the birthrate discrepancy between
millisecond pulsars and LMXBs disappears, indicating that LMXBs are
probably sufficient to produce the observed millisecond pulsars.

\acknowledgements

We thank the Penn State team, particularly P. Broos, for the
development and support of the ACIS\_EXTRACT software.  
We thank D. Lloyd for the use of his neutron
star atmosphere models.    
We also thank P. Jonker and R. Wijnands for useful discussions and
insightful comments, and the anonymous referee for a useful report.  
C.~O.~H. acknowledges the support of \Chandra grant GO2-3059A, and of the
Lindheimer fund at Northwestern University. 

\bibliography{src_ref_list}
\bibliographystyle{apj}


\clearpage


\clearpage

\clearpage

\clearpage

\begin{deluxetable}{lccccr}
\tabletypesize{\footnotesize}
\tablewidth{4.8truein}
\tablecaption{\textbf{Summary of \Chandra Observations}}
\tablehead{
\colhead{Seq,OBS\_ID} & \colhead{Start Time} & \colhead{Exposure} &
\colhead{Aimpoint} & \colhead{Frametime} & \colhead{CCDs}
}
\startdata
300003,078 & 2000 Mar 16 07:18:30  &  3875  & ACIS-I & 0.94 & 1/4 \\
300028,953 & 2000 Mar 16 08:39:44  & 31421  & ACIS-I & 3.24 & 6 \\
300029,954 & 2000 Mar 16 18:03:03  &   845  & ACIS-I & 0.54 & 1/8 \\
300030,955 & 2000 Mar 16 18:33:03  & 31354  & ACIS-I & 3.24 & 6 \\
300031,956 & 2000 Mar 17 03:56:23  &  4656  & ACIS-I & 0.94 & 1/4 \\
400215,2735 & 2002 Sep 29 16:59:00 & 65237 & ACIS-S & 3.14 & 5 \\
400215,3384 & 2002 Sep 30 11:38:22 &  5307 & ACIS-S & 0.84 & 1/4 \\
400216,2736 & 2002 Sep 30 13:25:32 & 65243 & ACIS-S & 3.14 & 5 \\
400216,3385 & 2002 Oct 01 08:13:32 &  5307 & ACIS-S & 0.84 & 1/4 \\
400217,2737 & 2002 Oct 02 18:51:10 & 65243 & ACIS-S & 3.14 & 5 \\
400217,3386 & 2002 Oct 03 13:38:21 &  5545 & ACIS-S & 0.84 & 1/4 \\
400218,2738 & 2002 Oct 11 01:42:59 & 68771 & ACIS-S & 3.14 & 5 \\
400218,3387 & 2002 Oct 11 21:23:12 &  5735 & ACIS-S & 0.84 & 1/4 \\
\enddata
\tablecomments{Times in seconds.  Subarrays are indicated by
  fractional numbers of CCDs. 
} \label{tab:obs}
\end{deluxetable}

\clearpage
\thispagestyle{empty}
\begin{landscape}
  \begin{deluxetable}{lccccccccr}
    \tabletypesize{\scriptsize}
    \tablewidth{8.4truein}
    \tablecaption{\textbf{X-ray Spectral Model Parameters}}
    \tablehead{
      \colhead{\textbf{Obs}} & \colhead{$F_{\rm X,abs}$(0.5-10 keV)} 
& \colhead{$N_H$} & \colhead{kT} & \colhead{R} &
\colhead{$L_{X,NS}$(0.5-10 keV)} & \colhead{$L_{bol,NS}$} &
\colhead{$\Gamma$} & 
\colhead{$L_{X,PL}$(0.5-10 keV)} & \colhead{$\chi^2_{\nu}$/DoF(nhp)} \\ 
 & ($10^{-14}$ ergs s$^{-1}$) & ($10^{20}$ cm$^{-2}$) & (eV) & (km) & ($10^{31}$ ergs s$^{-1}$) &
($10^{31}$ ergs s$^{-1}$) & & ($10^{31}$ ergs s$^{-1}$) & \\
}             
    \startdata
    \tableline
    \multicolumn{10}{c}{\bf W17, 2000 and 2002 data} \\
    \tableline
    - & $1.5^{+0.3}_{-0.4}$ & $4.1^{+1.7}_{-1.3}$ & $46^{+11}_{-12}$  &
$15.0^{+15}_{-5.3}$ & $1.5$ & $7.3$ & $1.9^{+0.4}_{-0.3}$ &
$3.3$ & 1.28/43(10\%) \\
    \tableline
    \multicolumn{10}{c}{\bf W37, 2000 and 2002 data} \\
    \tableline
    - & -$^a$  & -$^a$ & $82^{+10}_{-9}$ & $12.3^{+5.8}_{-3.5}$ & 26 & 53 & 3.0$^{+4.8}_{-0.9}$ & 0.56 & 1.24/57(11\%) \\
    \tableline
    \multicolumn{10}{c}{\bf X4, $N_H$ and H-atmosphere fixed, power-law varies} \\
    \tableline
    2735 high  & 3.7$^{+1.3}_{-1.3}$ & $5.0^{+0.7}_{-1.4}$ & $53^{+13}_{-8}$
& $10.8^{+7.5}_{-4.7}$ & 2.0 & 7.3 & $2.4^{+0.3}_{-0.3}$ & 9.8 & 1.30/38(11\%) \\
    2735 low   & 1.3$^{+0.5}_{-0.7}$ & - & - & - &  - &  - &
$3.1^{+0.5}_{-0.7}$ & 2.5 & - \\ 
    Other 2002 & 2.0$^{+0.4}_{-0.9}$ & - & - & - &  - &  - &
$2.3^{+0.4}_{-0.3}$ & 4.6 & - \\ 
    2000       & 1.3$^{+0.7}_{-0.8}$ & - & - & - &  - &  - &
$2.4^{+1.0}_{-1.0}$ & 2.4 & - \\ 
    \tableline
    \multicolumn{10}{c}{\bf X4, $N_H$ and power-law fixed, H-atmosphere varies} \\
    \tableline     
    2735 high  & $3.2^{+0.3}_{-2.0}$ & $3.9^{+0.6}_{-1.2}$ & $97^{+24}_{-19}$ & 4.1+2.9-1.7 & 6.6 & 12 & 2.1$^{+0.2}_{-0.4}$ & 3.8 & 1.35/38(7\%) \\
    2735 low   & $1.7^{+0.3}_{-1.1}$ & - & $44^{+12}_{-11}$ & $19^{+22}_{-9}$     & 2.0 & 10  & - & - & - \\
    Other 2002 & $1.9^{+0.3}_{-0.7}$ & - & $66^{+12}_{-11}$ & $7.0^{+2.0}_{-2.6}$ & 2.7 & 7.1 & - & - & - \\
    2000       & $1.6^{+1.9}_{-1.1}$ & - & $28^{+32}_{-1,b}$ & $70^{+26}_{-62}$    & 1.5 & 24  & - & - & - \\
    \enddata
    \tablecomments{ Spectral fits to W37, W17 and separate parts of X4 data,
using \citet{Lloyd03} hydrogen-atmosphere neutron star model plus a 
power-law model.  For X4, some parameters are held fixed between
observations, while others are allowed to vary.  All errors are 90\%
confidence limits.  Distance of 
4.85 kpc is assumed.  Flux measurements are absorbed (no correction
for $N_H$), while $L_X$ and $L_{bol}$ are unabsorbed.  
Neutron star H-atmosphere radius and temperature for assumed
grav. redshift of 0.306, implying 10 km, 1.4 \Msun\ NS; this 
tests for consistency with the standard model.  \\
$^a$ These parameters vary among W37's observations, and are presented in
Table~\ref{tab:add}. \\
$^b$ This parameter reached the hard limit of the model.
}\label{tab:spec}
  \end{deluxetable}
  \clearpage
\end{landscape}

\begin{deluxetable}{lcccr}
\tabletypesize{\footnotesize}
\tablewidth{5.0truein}
\tablecaption{\textbf{Additional Parameters for W37}}
\tablehead{
\colhead{\textbf{Obs.}} & \colhead{Cts} & \colhead{$F_X$, (0.5-2.5 keV)}  &
\colhead{$N_H$} & \colhead{Intervals of spectral extractions} \\ 
 & & ($10^{-14}$ ergs s$^{-1}$)  & ($10^{20}$ cm$^{-2}$) & ($10^3$ s)  
}              
\startdata
2735,H & 618 & $7.4^{+0.8}_{-0.5}$  & $5.9^{+3.2}_{-1.9}$ & 29.006:43.006, 53.006:66.102 \\
2735,L & 338 & $3.2^{+0.9}_{-0.3}$  & $31^{+5}_{-5}$   & 0:29.006, 43.006:53.006 \\
2736   & 40 & $0.2^{+0.8}_{-0.1}$ & $724^{+1e6,a}_{-371}$ & 74.104:140.209  \\
2737,H & 94 & $2.7^{+0.8}_{-0.4}$  & $38^{+9}_{-7}$   & 266.385:278.006, 287.506:289.506  \\
2737,L & 87 & $0.7^{+0.7}_{-0.2}$ & $146^{+29}_{-20}$ & 278.006:287.506, 289.506:332.490  \\
2738,H & 29 & $1.5^{+1.2}_{-0.3}$  & $67^{+25}_{-14}$ & 1038.010:1045.010  \\
2738,L & 71 & $0.5^{+0.7}_{-0.2}$  & $201^{+48}_{-32}$ & 980.879:1038.010, 1045.010:1051.550  \\
2000   & 46 & $0.4^{+0.7}_{-0.1}$  & $248^{+76}_{-43}$ & -  \\
\enddata
\tablecomments{ Spectral fits to various parts of W37 data,
using \citet{Lloyd03} hydrogen-atmosphere neutron star model and
power-law model (as in Table~\ref{tab:spec}), with $N_H$ the only variable
parameter. Times are barycentered
and should be added to reference time 149706994.3 (seconds after MJD 50814.0).  All errors
are 90\% confidence limits.  Distance of  
4.85 kpc is assumed.    \\
$^a$ This parameter reached the hard limit of the model.
}\label{tab:add}
\end{deluxetable}

\end{document}